\documentclass[12pt]{article}
\usepackage{enumerate}
\usepackage{amsmath,amssymb}
\usepackage{epsfig,euscript,array}
\usepackage{axodraw}
\usepackage{subfigure}
\usepackage{color}
\usepackage[sort&compress,square]{natbib}
\usepackage{graphics,graphicx}

\makeatletter \@addtoreset{equation}{section} \makeatother

\definecolor{blue}{rgb}{0,0,1}
\definecolor{red}{rgb}{1,0,0}

\definecolor{green}{rgb}{0,1,0}

\newcommand{\case}[2]{{\scriptstyle \frac{#1}{#2}}}

\newcounter{multieqs}

\newcommand{\be}{\begin{equation}}
\newcommand{\ee}{\end{equation}}

\newcommand{\bm}[1]{\mbox{\boldmath $#1$}}

\def\bd{\begin{document}}
\def\ed{\end{document}}
\def\nn{\nonumber}
\def\bea{\begin{eqnarray}}
\def\eea{\end{eqnarray}}
\let\bm=\bibitem
\let\la=\label

\newcommand{\EQ}[1]{\begin{equation} #1 \end{equation}}
\newcommand{\AL}[1]{\begin{subequations}\begin{align} #1 \end{align}\end{subequations}}
\newcommand{\SP}[1]{\begin{equation}\begin{split} #1 \end{split}\end{equation}}
\newcommand{\ALAT}[2]{\begin{subequations}\begin{alignat}{#1} #2 \end{alignat}\end{subequations}}
\def\beqa{\begin{eqnarray}}
\def\eeqa{\end{eqnarray}}
\def\beq{\begin{equation}}
\def\eeq{\end{equation}}

\def\hf{{\textstyle \frac{1}{2}}}
\def\wbar{\bar w}
\def\mubar{\bar\mu}
\def\abar{\bar a}
\def\sigmabar{\bar\sigma}
\def\etabar{\bar\eta}
\def\zetabar{\bar\zeta}
\def\mubar{\bar\mu}
\def\nubar{\bar\nu}
\def\N{{\cal N}}
\def\sst{\scriptscriptstyle}
\def\thetabar{\bar\theta}
\def\Tr{{\rm Tr}}
\def\one{\mbox{1 \kern-.59em {\rm l}}}
 \def\Nh{\hat{N}}

\newlength{\myVSpace}% the height of the box
\newcommand\xstrut{\raisebox{-.5\myVSpace}% symmetric behaviour,
  {\rule{0pt}{\myVSpace}}%
}

\def\a{\alpha}      \def\da{{\dot\alpha}}
\def\b{\beta}       \def\db{{\dot\beta}}
\def\c{\gamma}  \def\G{\Gamma}  \def\cdt{\dot\gamma}
\def\d{\delta}  \def\D{\Delta}  \def\ddt{\dot\delta}
\def\e{\epsilon}        \def\vare{\varepsilon}
\def\f{\phi}    \def\F{\Phi}    \def\vvf{\f}
\def\h{\eta}
\def\k{\kappa}
\def\l{\lambda} \def\L{\Lambda}
\def\m{\mu} \def\n{\nu}
\def\o{\omega}
\def\p{\pi} \def\P{\Pi}
\def\r{\rho}
\def\s{\sigma}  \def\S{\Sigma}
\def\t{\tau}
\def\th{\theta} \def\Th{\Theta} \def\vth{\vartheta}
\def\X{\Xeta}
\def\z{\zeta}

\def\cA{{\cal A}} \def\cB{{\cal B}} \def\cC{{\cal C}}
\def\cD{{\cal D}} \def\cE{{\cal E}} \def\cF{{\cal F}}
\def\cG{{\cal G}} \def\cH{{\cal H}} \def\cI{{\cal I}}
\def\cJ{{\cal J}} \def\cK{{\cal K}} \def\cL{{\cal L}}
\def\cM{{\cal M}} \def\cN{{\cal N}} \def\cO{{\cal O}}
\def\cP{{\cal P}} \def\cQ{{\cal Q}} \def\cR{{\cal R}}
\def\cS{{\cal S}} \def\cT{{\cal T}} \def\cU{{\cal U}}
\def\cV{{\cal V}} \def\cW{{\cal W}} \def\cX{{\cal X}}
\def\cY{{\cal Y}} \def\cZ{{\cal Z}}

\def\ua{\underline{\alpha}}
\def\ub{\underline{\phantom{\alpha}}\!\!\!\beta}
\def\uc{\underline{\phantom{\alpha}}\!\!\!\gamma}
\def\um{\underline{\mu}}
\def\ud{\underline\delta}
\def\ue{\underline\epsilon}
\def\una{\underline a}\def\unA{\underline A}
\def\unb{\underline b}\def\unB{\underline B}
\def\unc{\underline c}\def\unC{\underline C}
\def\und{\underline d}\def\unD{\underline D}
\def\une{\underline e}\def\unE{\underline E}
\def\unf{\underline{\phantom{e}}\!\!\!\! f}\def\unF{\underline F}
\def\unm{\underline m}\def\unM{\underline M}
\def\unn{\underline n}\def\unN{\underline N}
\def\unp{\underline{\phantom{a}}\!\!\! p}\def\unP{\underline P}
\def\unq{\underline{\phantom{a}}\!\!\! q}
\def\unQ{\underline{\phantom{A}}\!\!\!\! Q}
\def\unH{\underline{H}}

\def\As {{A \hspace{-6.4pt} \slash}\;}
\def\bs {{b \hspace{-6.4pt} \slash}\;}
\def\Ds {{D \hspace{-6.4pt} \slash}\;}
\def\ds {{\del \hspace{-6.4pt} \slash}\;}
\def\ss {{\s \hspace{-6.4pt} \slash}\;}
\def\ks {{ k \hspace{-6.4pt} \slash}\;}
\def\ps {{p \hspace{-6.4pt} \slash}\;}
\def\pas {{{p_1} \hspace{-6.4pt} \slash}\;}
\def\pbs {{{p_2} \hspace{-6.4pt} \slash}\;}

\def\Fh{\hat{F}}
\def\Vh{\hat{V}}
\def\Xh{\hat{X}}
\def\ah{\hat{a}}
\def\xh{\hat{x}}
\def\yh{\hat{y}}
\def\ph{\hat{p}}
\def\xih{\hat{\xi}}

\def\psit{\tilde{\psi}}
\def\Psit{\tilde{\Psi}}
\def\tht{\tilde{\th}}

\def\At{\tilde{A}}
\def\Qt{\tilde{Q}}
\def\Rt{\tilde{R}}
\def\Nt{\tilde{N}}

\def\at{\tilde{a}}
\def\st{\tilde{s}}
\def\ft{\tilde{f}}
\def\pt{\tilde{p}}
\def\qt{\tilde{q}}
\def\vt{\tilde{v}}
\def\nt{\tilde{n}}

\def\delb{\bar{\partial}}
\def\bz{\bar{z}}
\def\bD{\bar{D}}
\def\bB{\bar{B}}

\def\bk{{\bf k}}
\def\bl{{\bf l}}
\def\bp{{\bf p}}
\def\bq{{\bf q}}
\def\br{{\bf r}}
\def\bx{{\bf x}}
\def\by{{\bf y}}
\def\bR{{\bf R}}
\def\bV{{\bf V}}

\def\d{\delta}\def\D{\Delta}\def\ddt{\dot\delta}

\def\pa{\partial} \def\del{\partial}
\def\xx{\times}
\def\uno{\mbox{1 \kern-.59em {\rm l}}}

\def\trp{^{\top}}
\def\inv{^{-1}}
\def\dag{{^{\dagger}}}
\def\pr{^{\prime}}

\def\rar{\rightarrow}
\def\lar{\leftarrow}
\def\lrar{\leftrightarrow}

\newcommand{\0}{\,\!}      %this is just NOTHING!
\def\one{1\!\!1\,\,}
\def\im{\imath}
\def\jm{\jmath}

\newcommand{\tr}{\mbox{tr}}
\newcommand{\slsh}[1]{/ \!\!\!\! #1}

\def\vac{|0\rangle}
\def\lvac{\langle 0|}

\def\hlf{\frac{1}{2}}
\def\ove#1{\frac{1}{#1}}

\def\Box{\square}
\def\ZZ{\mathbb{Z}}
\def\CC#1{({\bf #1})}
\def\bcomment#1{}
\def\bfhat#1{{\bf \hat{#1}}}
\def\VEV#1{\left\langle #1\right\rangle}
\def\vev#1{\langle{#1}\rangle}

\newcommand{\ex}[1]{{\rm e}^{#1}} \def\ii{{\rm i}}

\def\rr{{\rm r}} \def\rs{{\rm s}}\def\rv{{\rm v}}
\def\ri{{\rm i}}\def\rj{{\rm j}}

\newcommand{\lrbrk}[1]{\left(#1\right)}
\newcommand{\sfrac}[2]{{\textstyle\frac{#1}{#2}}}

\font\mybb=msbm10 at 12pt
\def\bb#1{\hbox{\mybb#1}}

\font\myBB=msbm10 at 18pt
\def\BB#1{\hbox{\myBB#1}}

\begin{document}
\noindent
\hspace*{13.5cm}DESY 06-111\\
%\hspace*{13.5cm}IPPP/06/nn\\
%\hspace*{13.5cm}DCPT/06/nn

\vspace{25pt}

\begin{center}

{\Large \bf  Vacuum Birefringence as a Probe of Planck Scale Noncommutativity\\
}

\vspace{30pt}

{\bf Steven A. Abel$^a$, Joerg Jaeckel$^b$, Valentin V.  Khoze$^a$
and Andreas Ringwald$^b$}

{\small \em {}$^a$Centre for Particle Theory, Durham University, Durham, DH1 3LE, UK\\
{}$^b$Deutsches Elektronen-Synchrotron DESY, Notkestrasse 85, D-22607  Hamburg, Germany\\

\vspace{10pt}

{\sffamily \tt s.a.abel@durham.ac.uk, jjaeckel@mail.desy.de, valya.khoze@durham.ac.uk,
andreas.ringwald@desy.de} }

\vspace{30pt} {\bf Abstract}
\end{center}
Because of ultraviolet/infrared (UV/IR) mixing, the low energy
physics of noncommutative gauge theories in the Moyal-Weyl
approach seems to depend crucially on the details of the
ultraviolet completion. However, motivated by recent string theory
analyses, we argue that their phenomenology with a very general
class of UV completions can be accurately modelled by a cutoff
close to the Planck scale. In the infrared the theory tends
continuously to the commutative gauge theory. If the photon
contains contributions from a trace-U(1), we would observe vacuum
birefringence, i.e. a polarisation dependent propagation speed, as
a residual effect of the noncommutativity. Constraints on this
effect require the noncommutativity scale to be close to the
Planck scale.

\noindent {} \setcounter{page}{0} \thispagestyle{empty}

\newpage

\section{Introduction}
Field theories on noncommuting spacetime,
\begin{equation}
\label{noncommutativity}
[x^\mu,x^\nu]=\ii\,\theta^{\mu\nu} \ ,
\end{equation}
receive a great deal of attention, not least because
they arise naturally in a particular,
Seiberg-Witten,
limit \cite{Seiberg:1999vs} of string theories,
see \cite{Douglas:2001ba,Szabo:2001kg,Chu:2005ev} for reviews. 
The corresponding effective field theories can be derived from
their commutative cousins by simply replacing
products of fields in the Lagrangian by Weyl-Moyal star
products\footnote{In the
following,
we will not consider a more indirect alternative approach
to noncommutativity which
uses the Seiberg-Witten map.
For comments on the relation
between the two approaches see
\cite{Alvarez-Gaume:2003mb,Jaeckel:2005wt,Abel:2006sp}).},
\begin{equation}
(\phi * \varphi) (x) \equiv \phi(x)\  e^{{\case{\ii}{2}}\theta^{\mu\nu}
\stackrel{\leftarrow}{\partial_\mu}
\stackrel{\rightarrow}{\partial_\nu}} \  \varphi(x) \ . \label{stardef}
\end{equation}
The parameter $\theta^{\mu\nu}$ then appears in the vertices of  
perturbation theory, and, since it has dimensions of
$mass^{-2}$, defines a second mass scale in the theory (besides the
string scale, $M_{s}$), the so-called noncommutativity scale, $M_{\rm NC}$.
A natural question to ask is {\em what
is the allowed range of $M_{\rm NC}$?}

In this paper we shall consider the above question in as general
a manner as possible using
UV cutoffs to mimic the effects of the UV complete theory.
As we go along we will compare our results with
the string realisation of noncommutative field theory,
namely strings in background magnetic ($B$) fields, which provide
a nice, divergence-free framework in which to examine noncommutativity.
Despite this obvious attraction of strings, most of the phenomenology
depends on very general properties of any UV completion (for example
that they should be divergence free, continuous and so on)
and we will see that they are well modelled by UV cutoffs.

Before proceeding, let us try to make our question a little more precise. We can estimate the
{\em possible} range of $M_{\rm NC}$ by invoking the notion of naturalness.
Considering the specific example of string theory for a moment,
pure noncommutative field theory is realised as
a special limit of open strings in a background $B^{\mu\nu}$ field,
in which closed string (i.e. gravitational) modes are decoupled,
leaving only open string interactions.
There is no potential for $B$ which as far as the string theory is concerned
is just a rather mild background,
so in principle $\theta^{\mu\nu}$ could be anything.
Nevertheless it seems reasonable to suppose that,
if nonperturbative
string 
physics fixes the value of $B$ to be nonzero, it
does so with vacuum expectation values (VEVs) of order one in string
units.\footnote{Note that $M_{\rm NC}\gg M_{\rm P}$
does not imply large VEVs for the $B$ fields.
In general, since $\theta\sim \frac{1}{const+B}B\frac{1}{const-B}$ (Lorentz indizes suppressed)
vanishing $B$ fields imply vanishing $\theta$ and therefore $M_{\rm{NC}}\to\infty$.}
A natural scale for $\theta$
would in that case be $\theta\sim M^{-2}_{s}$. Depending on the
scenario in question that still leaves open a huge possible range:
$M_{\rm{P}}^{-2}< \theta < M_{W}^{-2}$, with the Planck scale $M_{\rm P}$
and the weak scale $M_W$, the latter arising, 
for example, in large extra dimension scenarios.
What about other possible
UV completions? One role of any UV completion
would almost certainly be to describe quantum gravity. As
$\theta^{\mu\nu}$ is intimately involved in the properties of spacetime,
a mild assumption is that its ``natural'' value
would inevitably be determined by the only mass scale in
quantum gravity. Then one would assume that typically
$\theta \sim M_{\rm P}^{-2}$. As the string theory example shows, the
natural range of $\theta$ could be beefed-up by, for example,
large volumes of extra compact dimensions, but it is
difficult to see how much smaller but nonzero
values could arise very easily.
If there is noncommutativity, therefore, it is natural
that $\theta>M_{\rm P}^{-2}$,
or equivalently, $M_{\rm NC} <M_{\rm P}$. 

So our slightly refined question is, {\em can noncommutativity at energy scales as high as $M_{\rm P}$
lead to observable effects?} Surprisingly, the answer is yes.
As we shall see in this paper,
current observations and experiments already severely restrict
the range of allowed noncommutativity scales.
The reason for this lies in two interesting properties of noncommutative
field theories that need to be taken into account in the
construction of a viable
noncommutative standard model extension \cite{Khoze:2004zc,Chaichian:2001py}.
First, there are strong constraints on both the dynamics and the field content.
Only U($N$) gauge groups with matter fields in fundamental,
bifundamental and  adjoint representations are allowed
\cite{Matsubara:2000gr,Armoni:2000xr,Gracia-Bondia:2000pz,Terashima:2000xq,Chaichian:2001mu}
(for U(1) gauge groups charges are restricted to $\pm1,0$
\cite{Hayakawa:1999zf}). Second, as we will detail below, universality
does not hold and ultraviolet/infrared (UV/IR) mixing
occurs \cite{Minwalla:1999px,Matusis:2000jf,Khoze:2000sy,Hollowood:2001ng}.

In four {\em continuous}
dimensions (i.e. without any quantum gravity,
high energy cutoff or UV completion), noncommutative
models of this type seem to conflict badly
with experiment, as outlined in Ref.~\cite{Jaeckel:2005wt}. Either there are superfluous
massless degrees of freedom or a nonvanishing
(and Lorentz symmetry violating)
mass term for the photon. Since neither is observed this presents
a challenge for any attempt to construct a realistic extension of the
Standard Model based on a noncommutative space time.

However, the result of Ref.~\cite{Jaeckel:2005wt} was based on the
assumption that the gauge fields live on a \emph{continuous} four
dimensional space time.  In particular, it assumed that the four
dimensional noncommutative gauge theory is valid up to arbitrarily
large momentum scales. But if noncommutative gauge theories
are realised as low energy effective field theories
of some underlying theory such as string theory, this assumption
almost certainly requires modification. It is likely that noncommutative
field theory gets spectacularly modified at energy scales
approaching $M_{s}$. One possible avenue to explore then is
the possible effects of ``stringy features'' such as
additional compactified space dimensions.
These make the theory effectively higher dimensional at large momentum
scales which can be thought of as an intermediate stage
towards the string theory. Thanks to UV/IR mixing, the effects of
extra dimensions can be transmitted to the IR in the trace-U(1) photon
\cite{extdim}. 
Such effects can be analysed in a field theoretic
framework, and one may search for helpful properties such as the amelioration
of constraints on the noncommutativity scale due to, for example,
power law decoupling of the trace-U(1) photon in the IR~\cite{extdim}.

Despite the obvious attraction of
the field-theoretical approach in \cite{Jaeckel:2005wt,extdim}, 
the drawback
is that it is unable to describe effects arising from
physics above $M_s$. This regime
is described by the UV completion of the theory, whatever that may be.
Normally of course we would not have to worry about such a thing
because of universality: the influence of physics above a cutoff $\Lambda_{\rm UV}$
on the physics at a momentum scale $k$ is suppressed by
powers of $k/\Lambda_{\rm UV}$. 
If universality holds, a modification at very high
momentum scales \emph{cannot} modify the physics at much smaller
momentum scales. However, although ordinary renormalizable commutative theories
fall into this category, noncommutative theories do not,
because of UV/IR mixing,
\cite{Minwalla:1999px,Matusis:2000jf} and \cite{Khoze:2000sy,Hollowood:2001ng}.

The phenomenon of UV/IR mixing can be understood from a simple argument.
To account for the effects of noncommutativity, we are instructed
to replace ordinary products in our field theory by Weyl-Moyal
star products \eqref{stardef}.
This results in factors of $\exp(\ii\tilde{k}\cdot p)$ in the
(non-planar) loop integrals \cite{Filk},
where
$\tilde{k}^{\mu}=\theta^{\mu\nu}k_{\nu}$. Consider a
typical loop integral with massless particles in the loop,
\begin{equation}
\label{intnp}
\int \frac{d^4{p}}{(2\pi)^{4}}\frac{1}{p^2(p+k)^2}\exp(\ii\tilde{k}\cdot p).
\end{equation}
The oscillating phase
regularises the integral for large values of momentum $p$,
and the integral is dominated by regions where $\tilde{k}\cdot p \sim 1$,
or $|p|\sim M^{2}_{\textrm{NC}}/|k|$.
This value of $|p|$ is large when the external momentum $k$ is small. The large
momenta in the loop $p\sim M^{2}_{\textrm{NC}}/k$ indeed influence
the physics at small external momentum $k$. Now consider
the effect of heavy particles of mass $M$ in the loop.
When $|k|\gg M^{2}_{\textrm{NC}}/|p|$, the loop integral is
killed when $|p|\ll M$, and (broadly speaking) we may neglect
the contribution of heavy particles. But when
$|k|\ll M^{2}_{\textrm{NC}}/|p|$ the phase is irrelevant and the integral
receives contributions from large values of $p$, $|p| > M$.
In other words as we lower our external momentum $k$, we
access ever heavier modes in the loop.

In  general, therefore, a modification of the
noncommutative theory above a UV scale $\Lambda_{\rm UV}$ will indeed influence
physics below an infrared scale $\Lambda_{\rm{IR}}\sim
M^{2}_{\rm{NC}}/\Lambda_{\rm UV}$, as we will see in detail in
Sect. \ref{uvirmixing}.  The problematic mass term for the photon is
an effect of this UV/IR mixing. Hence, it seems plausible that this
problem can be treated with a UV modification of the theory.
As we already stated, our aim here is to determine
some 
general phenomenological features of noncommutative models
and test them against experimental constraints. At first sight
this looks like a hopeless task, since constraints corresponding to
the lowest energy scales (for example photon masses)
are influenced by 
the highest mass modes in the loop integrals. It looks as though
sooner or later we will run up against the UV completion of the theory,
at which point all hopes of generality will be lost. However, guided by
recent work in Ref.~\cite{Abel:2006hk},
we can determine some generic properties for
a large class of theories (cf. Sect.~\ref{mimic}).
Indeed with fairly mild assumptions (which are true for string theory),
the phenomenological effects of the UV completion, such
as for example the restoration of normal Wilsonian behaviour in the
deep IR, are well modelled by a simple UV cutoff.

As we have already mentioned,
there is an important difference between the set-up we use in the present paper and that of 
Refs.~\cite{Khoze:2004zc,Jaeckel:2005wt,extdim}
in the way we interpret the underlying noncommutative gauge theory from the
perspective of standard particle physics at low energies. The UV/IR mixing effects illustrated by the integral
\eqref{intnp} occur only in the trace-U(1) factors
of the ${\rm U}(N)$ gauge group(s); the SU($N$) degrees of freedom are free from the UV/IR mixing.
The results of the present paper show, that, in presence of a fundamental cutoff $\Lambda_{\rm UV}$,
the mixing of those U(1) gauge fields (affected by the UV/IR mixing) with the photon does not
cause severe problems such as generating a polarisation dependent photon mass. This differs
from the approach in
Refs.~\cite{Khoze:2004zc,Jaeckel:2005wt,extdim}, where $\Lambda_{\rm UV}=\infty$.
As will be
explained in the next Section,
through the UV/IR mixing, an ultimate UV cutoff induces an infrared scale
$\Lambda_{\rm{IR}}\sim M^{2}_{\textrm{NC}}/\Lambda_{\rm UV}$.
At energy scales below $\Lambda_{\rm{IR}}$, physics of all degrees of freedom
is governed by an ordinary commutative
low-energy effective theory
and the effects of the UV/IR mixing are very small.
We now place the Standard Model at energies below $\Lambda_{\rm{IR}}$.
This is different from the set-up in Refs.~\cite{Khoze:2004zc,Jaeckel:2005wt,extdim}
which in the language of this paper amounts to
$\Lambda_{\rm UV} =\infty$ and $\Lambda_{\rm{IR}}=0,$ thus implying that the Standard Model
was embedded into a noncommutative theory
at energies $E_{\rm SM}$ in the opposite region:
$\Lambda_{\rm{IR}} < E_{\rm SM} < M_{\textrm{NC}} < \Lambda_{\rm UV}$.

We will see in Sect. \ref{masscutoff} that the problem of the unwanted mass term for
the trace-U(1) photon caused by the UV/IR mixing \cite{Jaeckel:2005wt} softens considerably
at energies below $\Lambda_{\rm{IR}}$. Instead of a mass term one gets, at low momentum scales,
vacuum birefringence, i.e. a polarisation dependent propagation speed. 
If $M_{\textrm{NC}}$ is close enough to the cutoff scale
$\Lambda_{\rm UV}\sim M_{\textrm{P}}$, this vacuum birefringence can be pushed beyond the current
experimental limits.
Thereby, a window opens for $M_{\textrm{NC}}$
where noncommutativity is still allowed.
As experimental and observational sensitivity is likely to improve
in the near future, this provides an interesting probe for scales $M_{\textrm{NC}}$ very close to the Planck scale.

In the following we will concentrate on the case of a pure U(1) noncommutative gauge theory.
A pure U(1) gauge theory behaves qualitatively like the trace-U(1) factors of U($N$) theories and captures
all essential features of the UV/IR mixing.\footnote{We recall that the noncommutative U(1) is an interacting
theory, which is asymptotically free in the UV. Its commutative counterpart is of course a free
theory.}

The paper is organized as follows. In Sect.~\ref{uvirmixing}, we discuss the essential
features of UV/IR mixing and the running gauge coupling in the presence of an ultimate UV cutoff.
In the following Sect.~\ref{masscutoff}, we demonstrate how this leads to vacuum
birefringence. We discuss experimental and observational bounds on this effect and
the resulting constraints on the scale of noncommutativity $M_{\rm NC}$.
The validity of using a fundamental UV cutoff to simulate
the UV completion is outlined in Sect.~\ref{mimic}, where we make
a comparison with string theory, and use it as
evidence in support of our claim that the phenomenology outlined here is
very generic. Finally, we conclude in Sect.~\ref{conclusions}.

\section{UV/IR mixing 
in presence of a finite UV cutoff}\label{uvirmixing}
In noncommutative gauge theories, Lorentz symmetry is explicitly broken since the matrix $\theta$ on
the right hand side of \eqref{noncommutativity} is a constant matrix to be specified
in a fixed reference frame.
This allows an additional transverse (gauge invariant) structure that might be present
in the polarisation tensor\footnote{Here, and in the following we will concentrate on the case of a noncommutative U(1) gauge group. The generalisation to U($N$) gauge groups is straightforward. All statements remain
valid, when applied to the trace-U(1) part of the gauge group. The SU($N$) part is unaffected
by noncommutativity, independent of the presence of a cutoff.},
\begin{equation}
\label{poltensor} \Pi_{\mu\nu}
= \Pi_1
(k^2,\tilde k^2)
\, \left( k^2 g_{\mu\nu} - k_\mu k_\nu \right) + \Pi_{2}
(k^2, \tilde k^2)\,
\frac{\tilde{k}_{\mu}\tilde{k}_{\nu}}{\tilde{k}^2} \quad\textrm{with}\quad  
\tilde{k}^{\mu}=\theta^{\mu\nu}k_{\nu}.
\end{equation}
The $\Pi_{1}$ part multiplies the ordinary transverse structure and is related to the gauge running coupling via
\cite{Khoze:2000sy} 
\begin{equation}
\frac{1}{g^2(k,\tilde{k})}=\frac{1}{g^{2}_{0}}+ \Pi_{1}(k,\tilde{k}).
\end{equation}
$\Pi_{2}$ is a new Lorentz symmetry violating structure
\cite{Minwalla:1999px,Matusis:2000jf}. 
In theories with exact supersymmetry (SUSY) 
it is absent \cite{Matusis:2000jf,Khoze:2000sy}.
Its size is therefore related to the SUSY breaking scale
\cite{Alvarez-Gaume:2003mb}. 

Performing a one loop calculation for the polarisation tensor one obtains \cite{Khoze:2000sy,Alvarez-Gaume:2003mb},
\begin{equation}
\label{pipi}
\Pi_{\mu\nu}(k)=\Pi_{\mu\nu}(k,l=0)-\Pi_{\mu\nu}(k,l=\tilde{k}),
\end{equation}
with
\begin{eqnarray}
\label{pipi2}
\Pi_{\mu\nu}(k,l)\!\!&=&\!\!2\sum_{j}\alpha_{j}\int \frac{d^4q}{(2\pi)^{4}}\bigg\{d(j)\left[
\frac{(2q+k)_{\mu}(2q+k)_{\nu}}{(q^2+m^{2}_{j})((q+k)^{2}+m^{2}_{j})}-\frac{2\delta_{\mu\nu}}{q^2+m^{2}_{j}}\right]
\\\nonumber
&&\quad\quad\quad\quad\quad\quad\quad\,\,\, +4C(j)\frac{k^{2}\delta_{\mu\nu}-k_{\mu}k_{\nu}}{(q^{2}+m^{2}_{j})((q+k)^{2}+m^{2}_{j})}\bigg\}
\exp\left(\ii q\cdot l\right),
\end{eqnarray}
where the coefficients $\alpha_{j}$, $d(j)$ and $C(j)$ are given in Tab. \ref{coefficients}.

\begin{table}[t]
\begin{center}
\begin{tabular}{|c|c|c|c|c|}
\hline  j=& scalar  & Weyl fermion & gauge boson  & ghost \\
\hline $\alpha_{j}$ & -1 & $\frac{1}{2}$ & $-\frac{1}{2}$ &1  \\
\hline  $C_j$ & 0 &  $\frac{1}{2}$& 2 & 0 \\
\hline  $d_j$ & 1 & 2 & 4 & 1 \\
\hline
\end{tabular}
\end{center}
\caption{Coefficients appearing in the evaluation of the loop diagrams.}
\label{coefficients}
\end{table}

As we already stated in the introduction we want to model the UV finiteness of an underlying theory by cutting off all fluctuations above a UV scale $\Lambda_{\rm UV}$.
One suitable way to do this is by introducing a factor of $\exp(-\frac{1}{\Lambda_{\rm UV}^{2}t^{2}})$ in the integral over the Schwinger time $t$. One obtains (s. \cite{Alvarez-Gaume:2003mb}),
\begin{eqnarray}
\Pi_{\mu\nu}(k)&=& {1 \over
4\pi^2}\left(k^2\delta_{\mu\nu}-k_{\mu}k_{\nu}\right) \nonumber \\
&  & \,\,\times \,\,\sum_{j}\alpha_{j}\int_{0}^{1}dx\,
\left[4C(j)-(1-2x)^2 d(j)\right]
\left[K_{0}\left({\sqrt{A_{j}}\over \Lambda_{\rm UV}}\right)
-K_{0}\left({\sqrt{A_{j}}\over \Lambda_{\rm eff}}\right)\right]
\nonumber\\
&+& {1\over (4\pi)^2}\tilde{k}_{\mu}\tilde{k}_{\nu}\,\Lambda_{\rm
eff}^2\sum_{j}\alpha_{j}d(j)
\int_{0}^{1}dx\,A_{j}K_{2}\left({\sqrt{A_{j}}\over
\Lambda_{\rm eff}}\right) \nonumber \\
&+& \delta_{\mu\nu}\left[\mbox{ gauge non-invariant term }\right],
\label{polarisation}
\end{eqnarray}
where
\begin{equation}
A_{j}=m^{2}_{j}+x(1-x)k^{2}
\end{equation}
and
\begin{equation}
\frac{1}{\Lambda^{2}_{\rm{eff}}}=\frac{1}{\Lambda_{\rm UV}^{2}}+\tilde{k}^{2}.
\end{equation}
We will neglect the gauge non-invariant terms in the following. They can be treated and eliminated
by using modified Ward-Takahashi identities \cite{Reuter:1993kw,Ellwanger:iz,Freire:2000bq}.

The employed regularisation cuts off the modes $p\gtrsim\Lambda_{\rm UV}$ in the loop integral in a smooth way.
Of course there are lots of different possibilities to do this. Since universality does not hold, 
different regularisations will, in principle, lead to different results. However, as long as we leave 
the qualitative feature ``all momenta $p\gtrsim \Lambda_{\rm UV}$ are cut off'' holds, we expect that 
the qualitative results we obtain remain true. For some details on other choices for the implementation 
of the cutoff, see Appendix \ref{regulator}.

Let us first concentrate on $\Pi_{1}$, i.e. the running gauge coupling,
and for the moment eliminate $\Pi_{2}$ by considering a
theory with unbroken supersymmetry. 

\begin{figure}
\begin{center}
\scalebox{0.95}[0.95]{
\begin{picture}(190,180)(40,0)
\includegraphics[width=9.5cm]{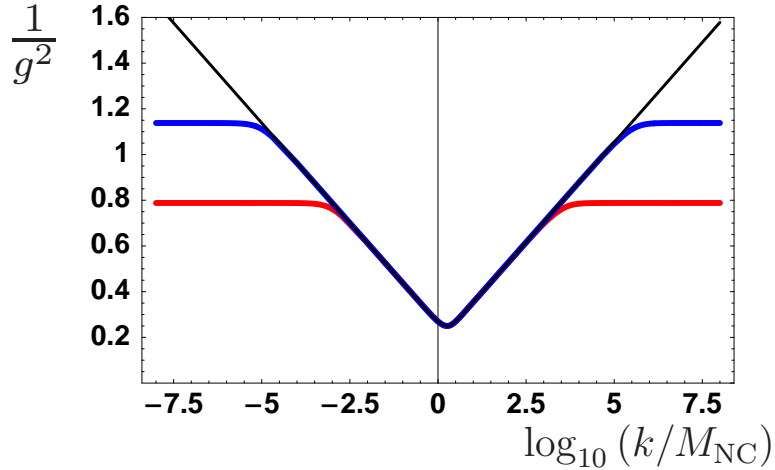}
\Text(-40,-10)[c]{\scalebox{1.4}[1.4]{$\log_{10}\left(k/M_{\textrm{NC}}\right)$}}
\Text(-285,150)[c]{\scalebox{2.0}[2.0]{$\frac{1}{g^2}$}}
\end{picture}
}
\end{center}
\caption{Running gauge coupling for a massless supersymmetric pure U(1)
gauge theory. The red, blue and black lines (bottom to top) are for
$\Lambda_{\rm UV}=1000\, M_{\textrm{NC}}, \,10^5 \,M_{\textrm{NC}},\,\infty \,M_{\textrm{NC}}$, respectively.
We have fixed the maximal gauge coupling to be $g^{2}_{\rm{max}}=4$. One can clearly see that for finite values
of the cutoff the running stops at $\sim \Lambda_{\rm UV}$ in the UV and at
$\Lambda_{\rm{IR}}\sim M^{2}_{\textrm{NC}}/\Lambda_{\rm UV}$
in the IR.} \label{gauge}
\end{figure}

In Fig. \ref{gauge}, we plot the running gauge coupling for various values of the cutoff $\Lambda_{\rm UV}$.
As expected the running stops at the UV scale $\Lambda_{\rm UV}$. In an ordinary commutative theory
we would expect no further changes. Here, however, we observe that the running stops, again, at an
infrared scale $\Lambda_{\rm{IR}}\sim M^{2}_{\textrm{NC}}/\Lambda_{\rm UV}$.
The running for $k< \Lambda_{\rm{IR}}$ vanishes up to threshold effects and is therefore essentially the same as that of a pure commutative U(1) gauge theory
(recalling that the $\beta$ function of a pure commutative U(1) gauge theory vanishes). It is easy to check that
a similar picture 
holds also for a more general matter content.
Stated differently, only in the range $\Lambda_{\rm{IR}}<k<\Lambda_{\rm UV}$
do
we observe a
truly noncommutative behavior of the running gauge coupling. Outside this range the behavior is
strongly affected by the presence of the UV cutoff.

So far we have rather sloppily been using the scale $M^{2}_{\rm{NC}}$.
Let us now give a more precise definition,
\begin{equation}
\label{absktilde}
|\tilde{k}| =\,  M^{-2}_{\rm{NC}}\,|k|,
\end{equation}
where $ M_{\rm{NC}}$ is the noncommutativity mass-scale.
Heuristically, $M^{-2}_{\rm{NC}} \sim |\theta|$ but it may depend on the direction.
E.g., for $\theta^{\mu\nu}$ in the canonical basis,
\begin{equation}
\theta^{\mu\nu} = \left(\begin{array}{cccc} 0 & \theta_1 & 0 & 0 \\
-\theta_1 & 0 & 0 &0 \cr 0 & 0 & 0 & \theta_2 \\
0 & 0 & -\theta_2 & 0 \end{array}\right)
,
\end{equation}
only when $\theta_1 \simeq \theta_2$
does one have
$M^{-2}_{\rm{NC}} = |\theta|.$ Otherwise the scale
$ M_{\rm{NC}}$ depends on $k_\mu,$
\begin{equation}
M^{-2}_{\rm{NC}} =\frac{ | \theta^{\mu\nu} k_{\nu} | }{|k|}=\,  |\theta_2|\,  \sqrt{1 +
\frac{\theta^{2}_{1}-\theta^{2}_{2}}{\theta^2_2} \frac{k_0^2+k_1^2}{k^2}}.
\end{equation}

\begin{figure}[!t]
\begin{center}
\subfigure{\scalebox{0.68}[0.68]{
\begin{picture}(190,260)(40,00)
\includegraphics[width=9.5cm]{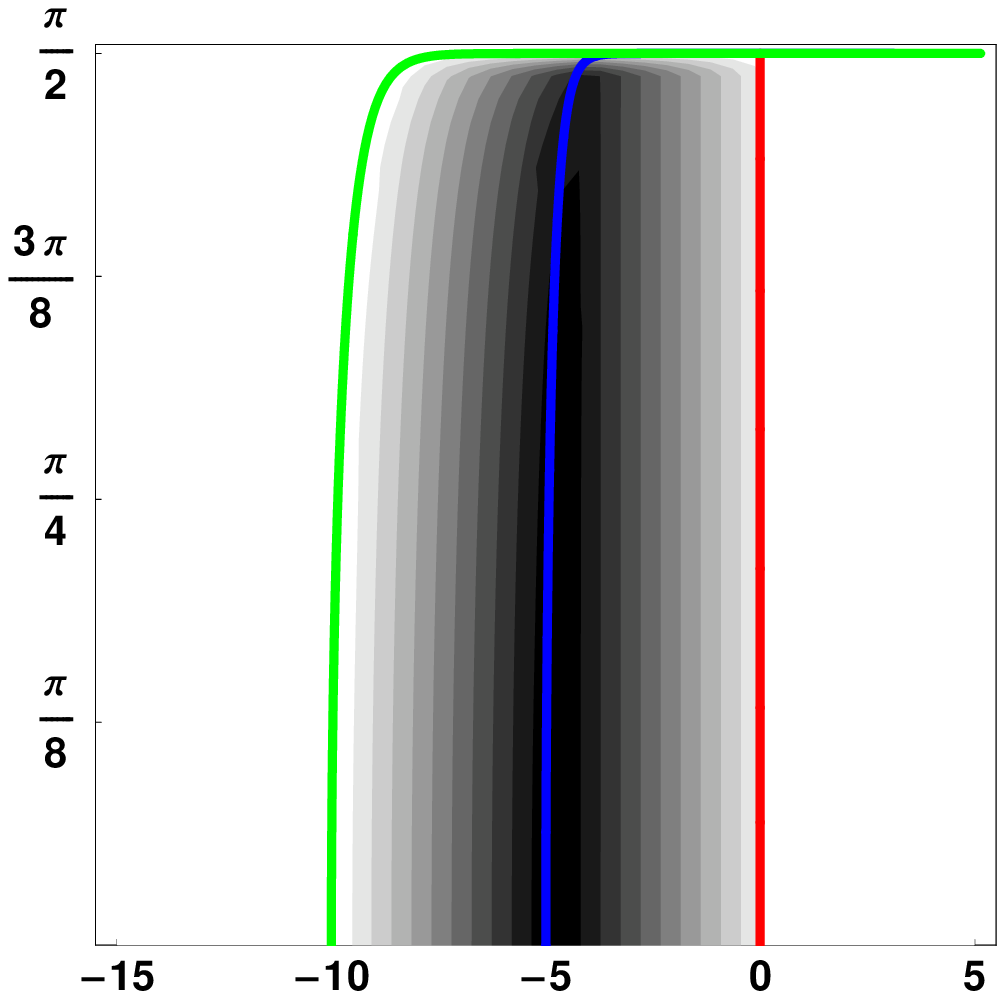}
\Text(-40,-10)[c]{\scalebox{1.4}[1.4]{$\log_{10}\left(k/\Lambda_{\rm UV}\right)$}}
\Text(-285,150)[c]{\scalebox{2.0}[2.0]{$\alpha$}}
\end{picture}
}}
\hspace{3cm}
\subfigure{\scalebox{0.68}[0.68]{
\begin{picture}(190,260)(40,0)
\includegraphics[width=9.5cm,height=9.4cm]{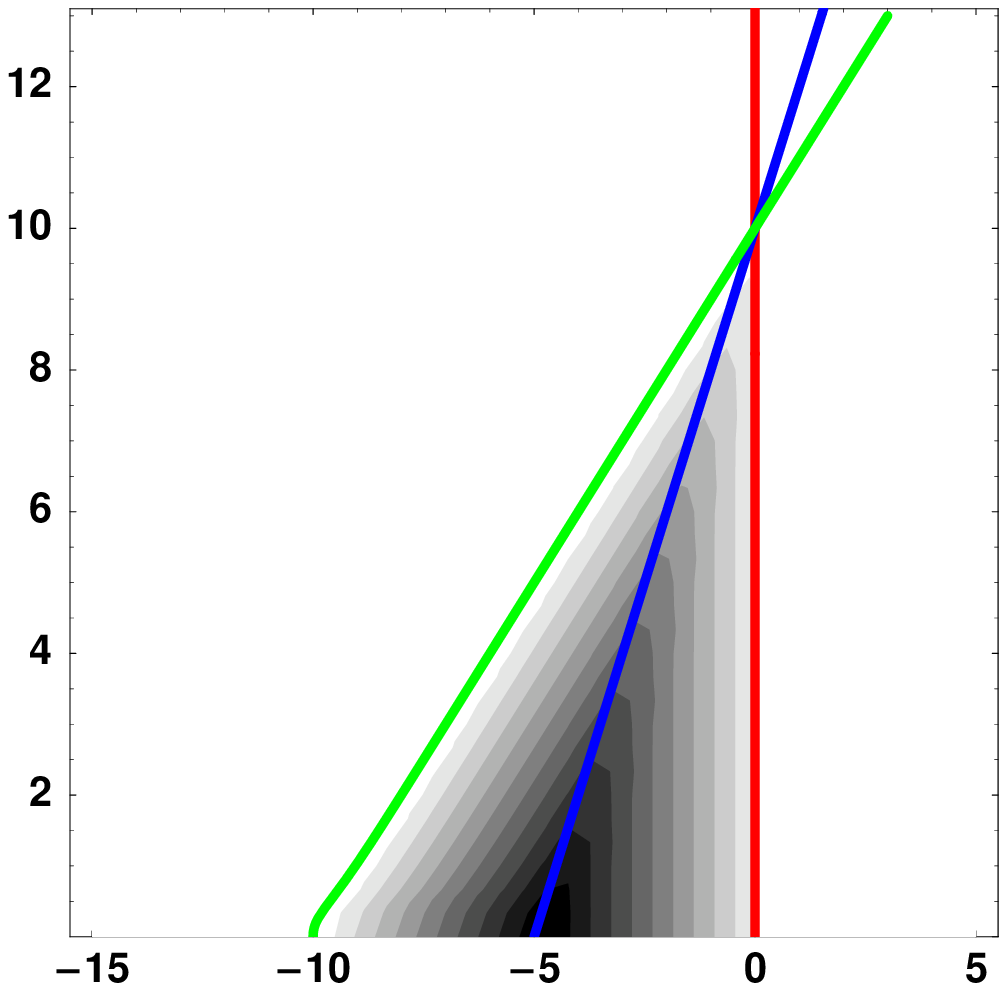}
\Text(-40,-10)[c]{\scalebox{1.4}[1.4]{$\log_{10}\left(k/\Lambda_{\rm UV}\right)$}}
\Text(-285,150)[c]{\scalebox{2.0}[2.0]{$\xi$}}
\end{picture}
}}
\end{center}
\caption{Depiction of the regions where the model behaves like a fully noncommutative theory (grey shaded area)
 and where the theory behaves more or less like the commutative theory (white). The green (left) line gives 
$\Lambda_{\rm{IR}}$, blue (middle) $M_{\rm{NC}}$ and red (right) $\Lambda_{\rm UV}$. 
In the left panel we have used 
$|\tilde{k}|=|k|\theta_{0}\cos(\alpha)$, $\theta_{0}=10^{10}/\Lambda_{\rm UV}^{2}$. In the right panel we 
use $\alpha=\arctan(10^{\xi}-1)$ to show that the lines for $\Lambda_{\rm{IR}}$ and $\Lambda_{\rm UV}$ intersect. The different grey shades also depict the deviation of $1/g^{2}$ from the UV value $1/g^{2}_{0}$ (lighter colors less deviation). This shows that in the IR and in the commutative directions 
($\alpha$ close to $\pi/2$) the coupling is given by the UV value. For a more general matter content the coupling will attain (up to threshold corrections) the same value as a purely commutative theory.} \label{region}
\end{figure}

If
for example one of the $\theta_{i}=0$ one can have a situation where $M_{\rm{NC}}\to\infty$. In general 
the truly noncommutative region $\Lambda_{\rm{IR}}<k<\Lambda_{\rm UV}$ will depend on the direction in 
momentum space. This is depicted in Fig. \ref{region}. While we have truly noncommutative behavior 
inside we have cutoff dominated nearly commutative behavior outside this region.

The crucial question is now in which region we perform experiments. If the noncommutativity scale is low and the cutoff
sufficiently
high, say $M_{\rm{NC}}\sim \rm{few} \,\rm{TeV}$, $\Lambda_{\rm UV} \sim 10^{18}\,\rm{GeV}$ we would live in the fully noncommutative region (shaded area in Fig. \ref{region}). However, this has already been excluded for a four dimensional theory \cite{Jaeckel:2005wt}. If we consider high scales for $M_{\rm{NC}}$,  say
$M_{\rm{NC}}\sim (10^{-3}-1)M_{\rm{P}}$, we find (using $\Lambda_{\rm UV}=M_{\rm{P}}$)
\begin{equation}
\Lambda_{\rm{IR}}\sim (10^{-6}-1)M_{\rm{P}}\gg k_{\rm{max}}\sim 1\,\rm{TeV}
,
\end{equation}
which is much bigger than the highest momentum transfer $k_{\rm{max}}$ that has been reached in experiments so far.
Therefore, with sufficiently high scale noncommutativity we expect to live in the nearly commutative region (white areas in Fig. \ref{region}) where the gauge coupling behaves like that of a commutative theory. Nevertheless, we will see in the next section that even in this region $\Pi_{2}$ leads to some possibly observable remnant effects.
One final remark concerning Fig. \ref{region} is that the deep IR and the far UV are
continuously connected. In this sense the UV and the deep IR are all part of
the ``UV phase''.

\section{Vacuum birefringence - a remnant effect from high scale noncommutativity}\label{masscutoff}
Let us now turn to the $\Pi_{2}$ part of the polarisation tensor. It, too, is affected by
the presence of a finite UV cutoff.
From Eq. \eqref{polarisation} we can easily see that it vanishes for a supersymmetric theory since
\begin{equation}
\sum_{j}\alpha_{j}d(j)=0
\end{equation}
for supersymmetric theories.
When supersymmetry is softly broken\footnote{Meaning that numbers of bosonic and fermionic
degrees of freedom of the theory still match.} 
we can easily derive the
following approximate expressions (for some additional details see Appendix \ref{regulator}),
\begin{eqnarray}
\label{approximate}
\Pi_{2}\!\!&=&\!\!D \Delta M^{2}_{\textrm{SUSY}},
\quad\quad\quad\,\,\,\,\,\textrm{for}\quad \frac{M^{2}_{\textrm{NC}}}{\Lambda_{\rm UV}}\ll k\ll \Delta M_{\textrm{SUSY}},
\\\nonumber
\Pi_{2}\!\!&=&\!\! D^{\prime} \Delta M^{2}_{\textrm{SUSY}}\Lambda_{\rm UV}^2\tilde{k}^2,
\quad\,\,\,\, \textrm{for}\quad k\ll\frac{M^{2}_{\textrm{NC}}}{\Lambda_{\rm UV}}, \,\,\,m^{2}_{j}\ll\Lambda_{\rm UV}^{2},
\end{eqnarray}
where $D,D^{\prime}$ are known constants and
\begin{equation}
\Delta M^{2}_{\textrm{SUSY}}=\frac{1}{2}\sum_{b}M^{2}_{b}-\sum_{f}M^{2}_{f}
\end{equation}
is the (super-)trace of the mass matrix.
Following the arguments given in \cite{Jaeckel:2005wt} we can now solve the equations of
motion for the photon, 
\begin{equation}
\label{eom}
\Pi^{\mu\nu}(k)A_{\nu}(k)=0.
\end{equation}

For concreteness, we now specify the noncommutativity,
\begin{equation}
\theta^{13}=-\theta^{31}=\theta :=\frac{1}{M^{2}_{\textrm{NC}}},
\end{equation}
and all other components of $\theta^{\mu\nu}$ vanishing (in the 3-direction, this use
of $M_{\rm NC}$ coincides with our direction dependent definition \eqref{absktilde}).
The photon flies in the three direction, 
\begin{equation}
k^{\mu}=(k^{0},0,0,k^{3}).
\end{equation}
Due to gauge invariance, only the two transverse polarisations are physical.
They have the polarisation vectors
\begin{equation}
A^{\mu}_{1}=(0,1,0,0),\quad A^{\mu}_{2}=(0,0,1,0).
\end{equation}
Inserting into Eq. \eqref{eom} we find
\begin{eqnarray}
(\Pi_{1}k^2-\Pi_{2})A^{\mu}_{1}\!\!&=&\!\!0,
\\\nonumber
\Pi_{1}k^2 A^{\mu}_{2}\!\!&=&\!\!0.
\end{eqnarray}
The photon polarized along $A^{\mu}_{2}$ obviously behaves like an ordinary massless photon. However, in the $A^{\mu}_{1}$
direction we observe new and interesting effects. To study these in more detail let us
now insert the approximate expressions \eqref{approximate}. For the $A^{\mu}_{1}$ polarisation we obtain
the following dispersion relations,
\begin{eqnarray}
\label{limit1}
&&\!\!\!\!k^2-D \frac{\Delta M^{2}_{\textrm{SUSY}}}{\Pi_{1}}=0,
\quad\quad\quad\quad\quad\quad\quad\,\,\,\, \textrm{for}\quad\Lambda_{\rm{IR}}=\frac{M^{2}_{\textrm{NC}}}{\Lambda_{\rm UV}}\ll k\ll \Delta M_{\textrm{SUSY}},
\\
\label{limit2}
&&\!\!\!\!k^2+D^{\prime} \frac{1}{\Pi_{1}} \frac{\Delta M^{2}_{\textrm{SUSY}}\Lambda_{\rm UV}^{2}}{M^{4}_{\textrm{NC}}} (k^3)^2=0,
\quad\quad\quad\textrm{for}\quad k\ll\frac{M^{2}_{\textrm{NC}}}{\Lambda_{\rm UV}}=\Lambda_{\rm{IR}}.
\end{eqnarray}

Equation \eqref{limit1} yields a Lorentz symmetry violating mass term of the order of 
$\Delta M^{2}_{\textrm{SUSY}}$
that was already discussed in detail in \cite{Jaeckel:2005wt}. Without cutoff, 
i.e. in the limit $\Lambda_{\rm UV}\to\infty$, 
this mass term persists down to $k\to 0$, thereby excluding any chance that this can be the photon observed
in nature. In presence of the cutoff, Eq. \eqref{limit1} is only applicable
for $k\gg \Lambda_{\rm{IR}}$.
Masslessness of the photon is well tested up to at
least $1\,\textrm{GeV}$. Using $M_{\textrm{P}}\sim \Lambda_{\rm UV} = 10^{18} \,\textrm{GeV}$, 
this gives us a conservative lower bound of $M_{\textrm{NC}}>10^{9}\,\textrm{GeV}$. Nevertheless,
this opens a rather large window of opportunity compared to the $\Lambda_{\rm UV}\to\infty$ case where
all $M_{\textrm{NC}}<M_{\rm{P}}$ are excluded.

For small photon momentum, Eq. \eqref{limit2} applies 
(recall from our discussion at the end of Sect. \ref{uvirmixing} that we actually expect to live in this limit). To understand \eqref{limit2} better, let us restore
the light speed $c$ in our equations and use $k^{0}=\omega$ for the frequency of the wave,
\begin{eqnarray}
\label{speed}
\omega^2-c^2\left(\frac{1}{1+\Delta n}\right)^2 (k^3)^2=0,
\end{eqnarray}
with
\begin{eqnarray}
\label{deltan}
\Delta n\!\!&\approx&\!\!
\frac{D^{\prime}}{2} \frac{1}{\Pi_{1}} \frac{\Delta M^{2}_{\textrm{SUSY}}\Lambda_{\rm UV}^{2}}{M^{4}_{\textrm{NC}}}
\\\nonumber
\!\!&=&\!\! 10^{-34}\left(\frac{D^{\prime}/2\Pi_{1}}{10^{-4}}\right)\left(\frac{\Delta M_{\rm{SUSY}}}{10^{3}\,\rm{GeV}}\right)^{2}\left(\frac{\Lambda_{\rm{UV}}}{10^{18}\,\rm{GeV}}\right)^{2}
\left(\frac{M_{\rm{NC}}}{10^{18}\,\rm{GeV}}\right)^{-4}\ll 1.
\end{eqnarray}
Here, we have combined the regularisation dependent loop factor
$D^{\prime}={\mathcal{O}}(1/4\pi^{2})$ and the field content
dependent factor $\Pi_{1}={\mathcal{O}}(10-100)$ to parameterise
the model dependence. 

From Eq. \eqref{speed} we can see that the photon $A^{\mu}_{1}$ propagates with a
speed\footnote{Note that $\Delta n<0$ is not inconsistent. Since Lorentz symmetry is explicitly broken,
 propagation with speeds $>c$ is, in principle, possible.} $\approx c(1-\Delta n)$. Since the $A^{\mu}_{1}$ photon propagates with $c$ we observe
birefringence, i.e. different polarisations propagate with different speed.

Although $\Delta n$ seems to be quite small we should compare this to the current experimental sensitivity.
In Ref.~\cite{Kostelecky:2002hh}, a study of all possible dimension four Lorentz violating operators in electrodynamics was
conducted and constraints derived.
The most general dimensions four Lagrangian which is gauge and CPT invariant, but violates Lorentz symmetry is,
\begin{equation}
\label{general}
{\mathcal{L}}_{\textrm{general}}=-\frac{1}{4}F^{\mu\nu}F_{\mu\nu}-\frac{1}{4}(k_{F})_{\mu\nu\alpha\beta}F^{\mu\nu}F^{\alpha\beta}.
\end{equation}
Comparing the propagator derived from Eq. \eqref{general} with Eq. \eqref{poltensor} we find
\begin{equation}
(k_{F})_{\mu\nu\alpha\beta}=\frac{D^{\prime}}{2}\frac{1}{\Pi_{1}}\Delta M^{2}_{\textrm{SUSY}}\Lambda_{\rm UV}^{2}\theta_{\mu\nu}\theta_{\alpha\beta}.
\end{equation}
In Ref. \cite{Kostelecky:2002hh}, the coefficients of $k_{F}$ have been constrained using various methods. For laboratory measurements, their estimate translates to 
\begin{equation}
|\Delta n_{\textrm{lab}}|\lesssim 10^{-14}-10^{-10},
\end{equation}
depending on the pattern of the noncommutativity.
Astrophysical observations already provide a much tighter bound of
\begin{equation}
|\Delta n_{\textrm{astro}}|\lesssim 10^{-16},
\end{equation}
while the strongest constraints come from observations of objects at cosmological distances
(see also \cite{Kostelecky:2001mb,Kostelecky:2006ta}), 
\begin{equation}
|\Delta n_{\textrm{cosmo}}|\lesssim  10^{-37}-10^{-32}.
\end{equation}

\begin{figure}
\begin{center}
\scalebox{0.95}[0.95]{
\begin{picture}(190,180)(40,0)
\includegraphics[width=9.5cm]{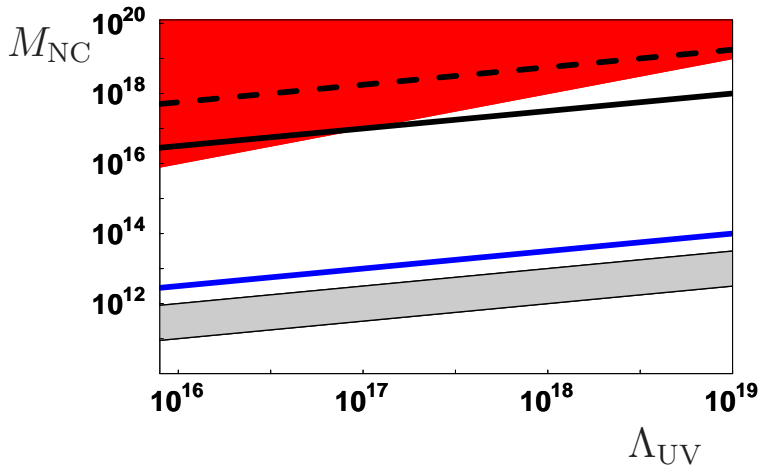}
\Text(-40,-10)[c]{\scalebox{1.4}[1.4]{$\Lambda_{\rm{UV}}$}}
\Text(-285,150)[c]{\scalebox{1.4}[1.4]{$M_{\rm{NC}}$}}
\end{picture}
}
\end{center}
\caption{Bounds on the scale of noncommutativity $M_{\rm{NC}}$ in a four dimensional noncommutative theory with an ultimate UV cutoff $\Lambda_{\rm{UV}}$. The red area is excluded by the requirement that $M_{\rm{NC}}<\Lambda_{\rm{UV}}$. The other curves show lower limits on $M_{\rm{NC}}$ derived via Eq. \eqref{deltan} from bounds on vacuum birefringence. The grey band corresponds to estimates from lab measurements. The blue (lower) and black (upper) thick solid curves originate from observations at astrophysical and cosmological distances, respectively. The thick dashed line gives the most recent constraint from polarisation measurements of 
gamma ray bursts \cite{Kostelecky:2006ta}. 
We used $D^{\prime}/2\Pi_{1}=10^{-4}$.} \label{bounds}
\end{figure}

In Fig. \ref{bounds}, we show the lower limits on $M_{\rm{NC}}$ originating from these experimental and observational
upper limits on the birefringence of the vacuum.

\section{\label{mimic}Cutoffs as a mimic of UV physics}

After having found that an ultimate UV cutoff leads to interesting physics
in noncommutative gauge theories, let us now discuss why
such a cutoff provides a good approximation to the effect of UV completion.
As we stated in the Introduction, our evidence for this comes
from the understanding one gains from the string theory theory
realisation of noncommutativity.

In order to appreciate what happens in string theory,
consider what would happen in a more general field theory
containing a tower of massive modes of mass $m_i$. In Euclidean space, the
typical one loop diagrams would have a sum over the modes as follows
\begin{equation}
\label{intnpm}
I(\theta,k,\Lambda_{\rm UV})=
\sum_i \int \frac{d^4{p}}{(2\pi)^{4}}\frac{1}{(p^2+m_i^2)((p+k)^2+m_i^2)}\exp(\ii\tilde{k}\cdot p).
\end{equation}
Again we will go to the Schwinger parameterisation.
Using the identity
\beq
\frac{1}{A_1A_2}=\int_0^1dx \int_0^\infty {dt}\,{t}\,{\rm e}^{-t(xA_1+(1-x)A_2)},
\eeq
we may recast
\eqref{intnpm} as
\beq
\label{intnpms}
I(\theta,k,m_i)=\int_0^\infty {dt}\,{t}\,
 {\rm e}^{-\frac{\tilde{k}^2}{4t}} \sum_i \int \frac{d^4{p}}{(2\pi)^{4}}
\int_0^1 dx \, {\rm e}^{-t (p^2 +k^2 x(1-x)+m_i^2)}
,
\eeq
where we used $k\cdot\tilde{k}=0$. Very heuristically,
the way string theory works as a finite UV completion
is to arrange the masses
$m_i$ so that the integrand resums into functions with
modular properties that render the integral finite.
The additional modes that are required to do this
have masses of order the string scale, i.e. the typical
masses of the lowest lying extra modes is order $M_s$; we call them UV modes\footnote{Actually the
question of finiteness in the string theory is quite subtle in this context. It relies on consistency
conditions, namely tadpole cancellation. For more details see Ref.~\cite{Abel:2006hk}.}.

More generally, we can consider the class of theories
where the UV completion
yields a one-loop contribution of the form
\beq
I(\theta,k,\Lambda_{\rm UV})=\int_0^\infty {dT}\,{T}\,
f\left( \frac{{\scriptsize{\tilde{k}^2 \Lambda_{\rm UV}^2}} }{4T}\right)
Z\left( T,\frac{k}{\Lambda_{\rm UV}}\right),
\eeq
where we have rescaled to a dimensionless Schwinger parameter, $t=T/\Lambda_{\rm UV}^2$.
The function $f$ contains all the effects of noncommutativity, whereas $Z$
would also be present in a commutative theory. Since the commutative theory
should be finite, too,
$Z$ 
implements the UV finiteness of the integral.
This property is typically provided by the sum over an appropriate spectrum of massive
modes as indicated by the sum in Eq.~\eqref{intnpms}.
It is this connection which imbues $\Lambda_{\rm UV}$ with a physical meaning
as the scale at which the UV completion modifies the integrand.
Roughly, it corresponds to the typical mass of the lightest UV modes $m_i$.
How this works in a string theoretical setting with a non-zero $B$-field
has been shown in Ref.~\cite{Abel:2006hk}. There,
the role of $\Lambda_{\rm UV}$ is played by $1/\sqrt{\alpha'}$,
with the string tension $\alpha'$. Moreover,
in that case the form of the function $f$ is indeed given by
\beq
f={\rm e}^{-\frac{\tilde{k}^2 \Lambda_{\rm UV}^2 }{4T}},
\eeq
for all $T$ (cf. Eq.~\eqref{intnpms}). In general, we expect
this simple form only for small
loop-momenta $\ll \Lambda_{\rm UV}$,
corresponding to $T\gg 1$,
\beq
\lim_{T\gg 1}f = {\rm e}^{-\frac{\tilde{k}^2 \Lambda_{\rm UV}^2 }{4T}}.
\eeq

A crucial determinant of the behaviour is the interplay between
$f$ and $Z$. In order to describe this further, we will
define two properties of the UV complete theory that
we will consider to be necessary:
\begin{itemize}
\item {\em All couplings in the $k\rightarrow 0$ limit tend to the couplings
of the $\theta=0$ theory.}
\item {\em All physics in the $\theta \rightarrow 0$ limit tends continuously
to $\theta = 0$ physics.}
\end{itemize}
These we will take to be fundamental properties of a consistent
UV completion and are certainly true for both string theory
and the field theory with a cutoff.
As stated in the introduction, the non-zero $B$ field is a
mild background field that we can dial continuously to zero;
it would be very odd for there to be any sort of discontinuity at $B=0$.
At least the second of these assumptions is known to be false in
noncommutative field theory, but then of course that theory
does not provide a finite UV completion.
Both properties are obviously true for the one-loop contribution
above if the integral is finite and
uniformly convergent, and if the function $f$ is continuous.

Given these assumptions it is clear that the
behaviour of the theory is essentially determined by
whether it is the function
$f$ or $Z$ which is doing the regulating of the integral. If
the regularisation is controlled by $Z$, then the behaviour
must by continuity be identical to the commutative string theory. However as
we shall see more thoroughly at the end of this section,
for momenta in the intermediate range $\Lambda_{\rm IR}<k<\Lambda_{\rm UV}$,
the integral is regulated by $f$. This leads to the field theoretical
behaviour where the integral is effectively regulated by the
noncommutativity (cf. Sect.~\ref{uvirmixing}).

To examine the question of continuity further,
it is instructive to consider taking the $\theta\rightarrow 0$ limit
by scaling $\theta \rightarrow \lambda \theta$. The only place
$\theta$ appears is in $f$; we may redefine
$\Lambda_{\rm UV} \rightarrow\sqrt{\lambda} \Lambda_{\rm UV} $
and $k\rightarrow \sqrt{\lambda}k $. The net result is
\beq
I(\lambda \theta, k,\Lambda_{\rm UV}) =
I(\theta, \sqrt{\lambda} k, \sqrt{\lambda} \Lambda_{\rm UV})\, .
\eeq
This equation looks a bit peculiar but on inspection
it makes sense: it says that the effect of taking
the commutative limit is the same as lowering the
mass scales of all the modes of the UV completion
to zero and leaving $\theta$ untouched.
In other words, on the right hand side the threshold effects of an increasing number of the
additional UV modes are included whilst $\tilde{k}\cdot p \ll 1$, and in the
limit the one-loop
correction to the gauge coupling includes all the same contributions as the
commutative theory, thus proving the second property for gauge couplings,
namely that as $\theta\rightarrow 0$ they tend to the commutative ones.
Note that this last statement is only true because
of the assumed convergence of the integral in a UV finite theory.

However the second property we are demanding of our theory is actually a
stronger requirement than this;
noncommutativity introduces new operators where momenta are contracted with
$\theta$'s, and the second property says that they tend
to zero in the IR. This is especially surprising given that
in noncommutative field theory the very same operators are divergent
in the IR. A typical operator is precisely the contribution to the
vacuum polarisation tensor of the trace-U(1) photon,
\beq
\Pi_{\mu\nu}\supset \Pi_{2}
(k^2, \tilde k^2)\,
\frac{\tilde{k}_{\mu}\tilde{k}_{\nu}}{\tilde{k}^2}.
\eeq
$\Pi_2$ has dimensions of $mass^2 $, and in
a generic non-supersymmetric field theory (with no matching between
numbers of bosonic and fermionic degrees of freedom)
$\Pi_2 \sim 1/\tilde{k}^2$ \cite{Minwalla:1999px,Matusis:2000jf}. 
In the UV complete theory, this contribution is of the general form
\begin{equation}
\Pi_{\mu\nu}\supset \tilde{k}_{\mu}\tilde{k}_{\nu}
J(\theta,k,m_i)=
\tilde{k}_{\mu}\tilde{k}_{\nu}
\Lambda_{\rm UV}^4
\int_0^\infty {dT}\,{T}\,
g\left( \frac{{\scriptsize{\tilde{k}^2 \Lambda_{\rm UV}^2}} }{4T}\right)
Z\left( T,\frac{k}{\Lambda_{\rm UV}}\right),
\end{equation}
where $g$ is a function with the same properties as $f$. Importantly,
continuity removes the possibility that it could have any divergences
in $\tilde{k}^2$, and insists that in the limit $k\rightarrow 0$
the integral $J$ converges to
the value in the ``commutative theory'' which is of order unity.
In the deep IR, therefore, we must have
\beq
 \Pi_{2} \sim \tilde{k}^2 \Lambda_{\rm UV}^4
,
\eeq
rather than any sort of divergence. In presence of softly broken SUSY,
the $\Lambda_{\rm UV}^4$ is softened to $\Delta M_{\rm SUSY}^2 \Lambda_{\rm UV}^2$
(cf. Eq.~\eqref{approximate}).

At what momentum scale does this behaviour take over from the
usual noncommutative field theory behaviour? The
extra UV modes can only contribute in the integral when $ T< 1 $.
Outside this region, contributions from the UV modes in the Schwinger
integral are exponentially suppressed, and the one loop contributions
are approximately those of the UV divergent field theory. Here, in the
diagrams sensitive to the noncommutativity, the UV divergence is tamed by the functions
$f$, $g$, which act as a cutoff for modes with
$T < 4 \tilde{k}^2 \Lambda_{\rm UV}^2 $. When the second inequality
saturates the first, that is when
\beq \tilde{k}^2 > \frac{1}{4\Lambda_{\rm UV}^2}\, , \qquad {\rm i.e.} \quad
k^2 > \frac{M_{\rm NC}^4}{4\Lambda_{\rm UV}^2} \sim \Lambda_{\rm IR}^2,
\eeq
we never get contributions from UV modes
in the integral and the behaviour is entirely field theoretical.
On the other hand when $\tilde{k}^2$ is less than this value, there
is a region
$4 \tilde{k}^2 \Lambda_{\rm UV}^2<T<1 $ where the UV modes are contributing
significantly. In this regime, the integration tends to the values
that we deduced from the convergence and continuity properties of
the UV completion and approaches a finite value as $k\rightarrow 0$.
Thus we can define a ``deep-IR'' region,
\beq
|k|<\Lambda_{\rm IR} = \frac{M_{\rm NC}^2 }{\Lambda_{\rm UV}},
\eeq
in which one-loop integrals give approximately constant
contributions, and Wilsonian behaviour is restored.

All of these properties are true for string theory, and
by inspection, they are mimicked by the introduction of a
cutoff in the Schwinger integral,
thus justifying the approach
that we have taken in the previous sections.

\section{Conclusions}\label{conclusions}
Noncommutative gauge theories are not universal. Therefore, any discussion of low
energy effects requires the specification of the ultraviolet sector. In this work we considered
a noncommutative field theory model where the fluctuations with momenta larger than an ultraviolet
cutoff $\Lambda_{\rm UV}$ give an overall vanishing contribution. We argued that this
is a good approximation to a large class of more fundamental UV finite theories, which
includes string theory.

The presence of an ultraviolet cutoff $\Lambda_{\rm UV}$ induces an effective infrared scale
$\Lambda_{\textrm{IR}}\sim M^{2}_{\textrm{NC}}/\Lambda_{\rm UV}$ below which the running coupling
behaves up to threshold corrections like that of a commutative gauge
theory\footnote{This is in stark contrast to a situation where the noncommutative gauge theory
is assumed to be valid at all scales and no ultraviolet cutoff exists. There $\Lambda_{\textrm{IR}}=0$ and the
theory shows strong effects of noncommutativity at all scales. In such a situation a noncommutative
U(1) can never be the photon as demonstrated in \cite{Jaeckel:2005wt,Alvarez-Gaume:2003mb}.}.
Only in the range $\Lambda_{\rm{IR}}<k<\Lambda_{\rm UV}$
do we observe full noncommutative behavior.
However, for large noncommutativity scales $M_{\rm{NC}}\gtrsim 10^{11}$~GeV and a cutoff
$\Lambda_{\rm UV}\sim M_{\rm{P}}$ one easily finds that all known experiments are performed
in the nearly commutative region $k<\Lambda_{\rm{IR}}$.

If supersymmetry is broken, an additional Lorentz symmetry violating structure is present in
the polarisation tensor. For scales $k>\Lambda_{\rm{IR}}$ it leads to a mass term for the gauge
boson in accord with Refs.~\cite{Jaeckel:2005wt,Alvarez-Gaume:2003mb}.
However, below $\Lambda_{\textrm{IR}}$ the mass term turns into a modification of the
phase velocity
of plane wave solutions, leading to birefringence.
If the trace-U(1) gauge boson is to be interpreted as (part of)
the photon, a mass is not acceptable and
birefringence must
be smaller than the experimental limits. Using the most stringent limits from cosmological observations
one obtains a rather strong limit of $M_{\textrm{NC}}\gtrsim 0.1 M_{\textrm{P}}$. If we use the more
conservative astrophysical or laboratory limits the same argument yields only
$M_{\textrm{NC}}\gtrsim (10^{-7}-10^{-5})\ M_{\textrm{P}}$.
In this setting high precision measurements of the properties of light are a wonderful tool to test (nearly)
Planck scale physics.

\bigskip
\bigskip

\centerline{\bf Acknowledgements}

We are indebted to Chong-Sun Chu and 
Mark Goodsell for discussions. SAA would like to thank CERN for hospitality.
VVK thanks the Aspen Centre for Physics for hospitality and
acknowledges support of PPARC
through a Senior Fellowship.

\bigskip

\begin{appendix}
\section{\boldmath Low energy $\Pi_{2}$ for general regulators}\label{regulator}
The effects discussed in this paper originate from the introduction of a UV cutoff $\Lambda_{\rm UV}$.
Let us now check that the qualitative behavior is independent of the specific implementation of the cutoff,
i.e. the choice of the function that suppresses fluctuations with $k\gtrsim \Lambda_{\rm UV}$. In particular,
let us check that the form given in \eqref{approximate} is generic for $k\ll \Lambda_{\rm{IR}}$.

Since the first term in \eqref{pipi} does not contain $\theta$ it is obvious that only the second term
can contribute to $\Pi_{2}$. Similarly the term in the second line of Eq. \eqref{pipi2} can only contribute
at order ${\mathcal{O}}(k^2 \tilde{k}^{2})$. Collecting the remaining terms of Eq. \eqref{pipi2} which are
$\propto \tilde{k}_{\mu} \tilde{k}_{\nu}$ one finds,
\begin{equation}
\Pi_{2}=-2\tilde{k}^{2}\sum_{j}\alpha_{j}d(j)\int \frac{d^{4}q}{(2\pi)^{4}}\left[\frac{q^{2}}{P_{j}(q)}
-\frac{2}{d}\frac{q^4}{P^{2}_{j}(q)}\right]
\end{equation}
where $P_{j}(q)$ is the inverse propagator of the particle $j$. (In absence of a cutoff,
$P_{j}(q)=q^{2}+m^{2}_{j}$.)
In presence of any reasonable UV cutoff that acts for all particles identically (which, in particular,
respects SUSY) one obtains\footnote{A large class of different cutoff functions can be implemented by
using ERGE scheme regularisation (see, e.g., Appendix C of \cite{Jaeckel:2003uz}).},
\begin{equation}
\int \frac{d^{4}q}{(2\pi)^{4}}\left[\frac{q^{2}}{P_{j}(q)}
-\frac{2}{d}\frac{q^4}{P^{2}_{j}(q)}\right]=\Lambda_{\rm UV}^{4}
f\left(\frac{m^{2}_{j}}{\Lambda_{\rm UV}^{2}}\right).
\end{equation}
As long as
\begin{equation}
\sum_{j} \alpha_{j}d(j) f\left(\frac{m^{2}_{j}}{\Lambda_{\rm UV}^{2}}\right)\neq 0
,
\end{equation}
we will observe a birefringence effect as discussed in Sect. \ref{masscutoff}.
For $m^{2}_{j}\ll\Lambda_{\rm UV}^{2}$, we can further expand, 
\begin{equation}
f\left(\frac{m^{2}_{j}}{\Lambda_{\rm UV}^{2}}\right)=A+B\frac{m^{2}_{j}}{\Lambda_{\rm UV}^{2}}+\ldots.
\end{equation}
Remembering that $\sum_{j}\alpha_{j}d(j)=0$ we find
\begin{equation}
\Pi_{2}\sim \tilde{k}^{2}B\Delta M^{2}_{\rm{SUSY}}\Lambda_{\rm UV}^{2},
\end{equation}
as in \eqref{approximate}. We have checked for several regulators that $B\neq 0$. However, let us remark
that $B$ depends on the choice of the regulator.
\end{appendix}

\end{document}